\documentclass[a4paper,10pt]{article}
\usepackage[utf8]{inputenc}
\usepackage{url}
\usepackage{amsmath}
\usepackage{amssymb}
\usepackage{pdflscape}
\usepackage{rotating}
\usepackage{float}
\usepackage{xcolor}

\floatstyle{ruled}
\newfloat{algorithm}{tbp}{loa}
\providecommand{\algorithmname}{Algorithm}
\floatname{algorithm}{\protect\algorithmname}

\begin{document}
%\maketitle

\begin{center}
{\Large \textbf{A Simulation Approach for Change-Points on Phylogenetic Trees}}

\vspace{0.5cm}

BY ADAM PERSING$^{1}$, AJAY JASRA$^{2}$, ALEXANDROS BESKOS$^{1}$, DAVID BALDING$^{3}$, \& MARIA DE IORIO$^{1}$

{\footnotesize $^{1}$Department of Statistical Science, University College London, London, WC1E 7HB, UK.}\\
{\footnotesize E-Mail:\,}\texttt{\emph{\footnotesize a.persing@ucl.ac.uk, a.beskos@ucl.ac.uk, m.deiorio@ucl.ac.uk}}\\
{\footnotesize $^{2}$Department of Statistics \& Applied Probability, National University of Singapore, Singapore, 117546, SG.}\\
{\footnotesize E-Mail:\,}\texttt{\emph{\footnotesize staja@nus.edu.sg}}\\
{\footnotesize $^{3}$Genetics Institute, University College London, London, WC1E 6BT, UK.}\\
{\footnotesize E-Mail:\,}\texttt{\emph{\footnotesize d.balding@ucl.ac.uk}}

\vspace{0.35cm}

%\today
\end{center}

\begin{abstract}
We observe $n$ sequences at each of $m$ sites, and assume that they have evolved from an ancestral sequence that forms the root of a binary tree of known topology and branch lengths, but the sequence states at internal nodes are unknown. The topology of the tree and branch lengths are the same for all sites, but the parameters of the evolutionary model can vary over sites. We assume a piecewise constant model for these parameters, with an unknown number of change-points and hence a trans-dimensional parameter space over which we seek to perform Bayesian inference. We propose two novel ideas to deal with the computational challenges of such inference. Firstly, we approximate the model based on the time machine principle: the top nodes of the binary tree (near the root) are replaced by an approximation of the true distribution; as more nodes are removed from the top of the tree, the cost of computing the likelihood is reduced linearly in $n$. The approach introduces a bias, which we investigate empirically. Secondly, we develop a particle marginal Metropolis-Hastings (PMMH) algorithm, that employs a sequential Monte Carlo (SMC) sampler and can use the first idea. Our time-machine PMMH algorithm copes well with one of the bottle-necks of standard computational algorithms: the trans-dimensional nature of the posterior distribution. The algorithm is implemented on simulated and real data examples, and we empirically demonstrate its potential to outperform competing methods based on approximate Bayesian computation (ABC) techniques.
\end{abstract}

\section{Author Summary}

A phylogeny (or evolutionary tree) can explain the ancestral relationships among species based on similarities in their genetic sequences (e.g., DNA). A phylogenetic model will typically be parametrized by rates which can correspond to genetic mutations that occur within populations as they evolve over time. In many applications, it is reasonable to assume that genetic sequences share a common phylogenetic structure across all of their sites. However, to allow for greater modelling flexibility, it is often times desirable to allow for the evolutionary rate parameters to change across the length of the sequences. The main focus of this paper is estimating that rate variation across the sites. We consider a model that allows for neighbouring blocks of sites to be parametrized by different evolutionary rates (i.e., a change-point model), and we propose a novel computational scheme that enables a practitioner to fit this model to genetic sequences when the true number of change-points is unknown. Thus, our contribution is a methodology that infers both the number of distinct blocks along the length of the sequence and the values of the rates themselves. We empirically demonstrate the potential of our algorithm to outperform competing computational methods.

\section{Introduction}

A phylogeny (or evolutionary tree) is the most common structure employed to explain the evolutionary relationships among species (`taxa') based on similarities in their physical or (more usually) genetic characteristics. The branching pattern of the tree is usually referred to as its topology, and describes shared and independent periods of evolution of different taxa. The leaves of the tree correspond to observations on the taxa. In a rooted phylogenetic tree (Figure \ref{fig:AppliedInputTree} of Appendix \ref{sec:figures}), each internal node corresponds to a speciation event and represents the most recent common ancestor of all the taxa descended from that node. The length of the edges connecting the nodes (`branches') can be interpreted as the time between speciation events. 

The evolutionary analysis of molecular sequence variation is statistically challenging. Parsimony methods were among the first approaches for inferring phylogenies, but in recent years, great research effort has been devoted to likelihood-based methods, both in the frequentist \cite{fel_81} and Bayesian framework. 

DNA sequences occupy one of four states (A, C, G, T) at each site, and so specifying the likelihood function requires a model for how these change over time at each site. The simplest such model is the Jukes-Cantor, in which each state is substituted by any other state at the points of a homogeneous Poisson process. The Kimura model has a rate for transitions (A $\leftrightarrow$ G or C $\leftrightarrow$ T) that can differ from the rate for transversions (all other substitutions), see \cite[Chapter 13]{Felsenstein_2004}. Objects of inference can include the topology of the phylogenetic tree (here regarded as known), the relative branch lengths on the tree, and the substitution rates. 

Likelihood-based approaches usually assume that substitution rates are the same at all sites, so that the likelihood is obtained as a product across sites. However, variation in substitution rates along DNA sequences is well established \cite{hus_such}. This variation can be explained by variation in functional constraint across the genes encoded in the sequences. If the DNA sequence is from a coding region, natural selection may constrain variability at some sites more than others and therefore sites might exhibit different rates of evolution. Therefore, it is important to accommodate rate variation across sites in phylogenetic inference \cite{hus_hil,wake_1994}. One possibility is to estimate a different rate for each site \cite{swoff} but this is computationally demanding because of the large number of parameters, and the limited information per parameter leads to poor inferences. A better alternative is to assume that the rates at different sites are independent draws from a distribution, typically either a Gamma \cite{uzzell, nei} or a Log-Normal distribution \cite{olsen}. A more realistic model would assume that the rates are auto-correlated along the sequence. One possible solution is offered by Phylogenetic Hidden Markov (phylo-HMM) models, which allow for correlated rates between nearby sites \cite{yang_95,fel_chur}: the rate of evolution is modelled as a Markov process operating along the sequence and site specific rates are drawn from a finite set of values. The discrete number of ’rate categories’ represents one limitation of the phylo-HMM approach \cite{yang_94,siepel}, while another is the small number of taxa that can be accommodated with reasonable computational resources \cite{yang_93}. Alternatively, \cite{suchard} have developed a Bayesian multiple change-point model of rate variation along the DNA sequence, which assumes that sites are grouped into an unknown number of contiguous segments, each with possibly a different tree topology, as well different substitution rates and branch lengths. Several recent proposals involve finite mixtures of distributions to model heterogeneity across sites. In this case, the distribution of each site on the sequence is a mixture of multiple processes, each of which may have its own tree topology, branch lengths and substitution rates (e.g.~\cite{pagel,hus_such,loza}). \cite{wu} extend these ideas to infinite mixtures assuming a Dirichlet process prior.

The main focus of this paper is estimating evolutionary rate variation across sites assuming that the tree topology and branch lengths are known and the same at every site under analysis. The latter assumption is not very restrictive in most applications, which involve taxa that are separated by enough time that within-taxon coalescent variation is unimportant. Although substitution rates can vary along the sequence, they are assumed to be the same across all taxa at each site. Our proposed time-machine PMMH model is able to account for quantitative differences in rates of substitutions (e.g. sites with high rates versus sites with low rates), and can also allow different rates for different types of substitution (such as transitions and transversions).

%%%%%%%%%%%%%%%%%%%%%%%%%%%
Recently there has been a revival of interest in models which allow for variation in evolutionary rates due to an explosion in the availability of comparative sequence data, and consequent interest in comparative methods for the detection of functional elements (e.g., \cite{Boffelli_2003, rgsc_2004, mgsc_2002}). The model proposed in this paper is similar in spirit to early work on spatial variation of evolutionary rates ( e.g. \cite{yang_95,fel_chur}), which maintains a single consistent topology along the sequence but allows changes in evolutionary rates.  In this framework, given the rate at each site, each site is then assumed to evolve independently along the true phylogeny with that rate and the correlation between sites arises from the clustering of high and low rates at adjacent sites. However, most of these models allow for a small discrete number of ’rate categories’ into which sections of the sequences are sorted (\cite{yang_94, Siepel_2005b}) and many methods are limited to two-species comparisons as they become increasingly computationally expensive  when more species are included. Our proposed model overcomes both these difficulties, as the model  for evolutionary rates,  based on a multiple change point model,  is structurally simple and flexible so that the rates are not restricted to a finite set but estimated on-line. Moreover, the use of the “time machine” significantly speeds up computations.
%%%%%%%%%%%%%%%%%%%%%%%%%%%

\subsection{Specific Contributions}

Several negative mathematical results exist in the literature (e.g.~\cite{mossel}) for Markov chain Monte Carlo (MCMC) inference when the tree topology (and branch lengths) is unknown, and these have spurred the development of highly sophisticated Monte Carlo-based algorithms \cite{bouch}. Here, the tree topology and branch lengths are assumed to be known, but the position and number of change-points for the rates are unknown. In addition, as we will explain later, the cost of evaluating the likelihood will be an $\mathcal{O}(mp^2 n^2)$ operation ($p$ is the number of states at each site, $m$ the number of sites and $n$ the number of sequences). It follows that parameter inference requires expectations w.r.t.~a probability on a trans-dimensional state-space. Contructing efficient MCMC algorithms on trans-dimensional spaces is a notoriously challenging problem and the standard approach is to use reversible jump MCMC (RJMCMC) \cite{green}. Typically, and especially for our model, it is difficult to develop moves on the trans-dimensional state-space that are likely to be accepted, which is important here because likelihood computations are expensive. 

To deal with some of these inferential and computational issues, we propose:
\begin{itemize}
\item{To reduce the cost of computing the likelihood and assist the mixing of MCMC, through a likelihood approximation based on the time-machine principle \cite{Jasra_2011}.}
\item{To improve mixing compared to standard RJMCMC, by adapting an idea in \cite{Karigiannis_2013}, developing a particle marginal Metropolis-Hastings (PMMH) algorithm \cite{Andrieu_2010}, based on the sequential Monte Carlo (SMC) samplers method in \cite{DelMoral_2006}. This approach can benefit from the time-machine approach.}
\end{itemize}

In the time-machine approach, the unobserved sequence at the root, and possibly also other top-most nodes of the tree, are replaced with the stationary distribution of the substitution process. This can reduce the cost of computing the likelihood by a linear factor in $n$; this can allow larger datasets than would otherwise be manageable. The resulting estimates are biased, but in the examples below we find the bias to be smaller than for competitive methods. Indeed,
%investigate the bias and show that, under assumptions on the model, it can disappear at a geometric rate with regards to the %proximity of the cut to the top of the tree.
%the closer to the top of the tree, that that cut is made. 
we conjecture (and this is supported by empirical results) that our approach is competitive with other approximate methods, in particular approximate Bayesian computation (ABC); this latter method is often not appropriate for model selection problems as we describe in Section \ref{sec:abc}. An important point here is that the time-machine performs 
a `principled' approximation of the mathematical model. This is based on the general understanding that most of the information in the data is at the lower part of the tree, thus contrasting with an often ad-hoc selection of summary statistics in ABC approaches.

Our PMMH algorithm extends the idea in \cite{Karigiannis_2013}, both with regard to the methodology and the context of phylogenetic trees with change-points. The MCMC method will often generate (as we will explain in Section \ref{sec:Computationalmethods}) trans-dimensional proposals which are more likely to be accepted than standard RJMCMC algorithms. This is further aided by using the time-machine, which results in a less complex posterior with a faster likelihood evaluations. The combination of the above factors can lead to reliable, but biased, inference from moderate sized data-sets. As mentioned above, we expect the bias to be minimal relative to ABC methods.

This article is structured as follows. In Section \ref{sec:modelsandmethods} the model and methods are described; this includes our mathematical result on the bias. In Section \ref{sec:results} our empirical results are given. In Section \ref{sec:disc} we conclude the article with a discussion. The appendix provides further details of the methods.

\section{Model and Methods}\label{sec:modelsandmethods}
We first describe our change-point model and the associated Bayesian inference problem, then the time machine approximation, and the PMMH algorithm. The end of this section then briefly discusses some competing ABC methods that can also be used to perform Bayesian inference, but we make a case against using such algorithms in this context. Throughout the article, given a vector $(x_1,\dots,x_n)$ we define $x_{k:l}:=(x_{k},x_{k+1},\dots,x_l)$, $k\leq l\le n$; also, we use the notation $[k]=\{1,2,\ldots,k\}$.

\subsection{Phylogenetic Model}
We observe $n$ sequences of length $m$, such that each observation is $x_{ij}\in\{1,\dots,p\}$ with $1\le i\le n$, 
$1\le j\le m$. Similar to as in \cite{YussanneMa_2008}, it is assumed that the data originate from a rooted binary tree \cite[Chapter 1]{Felsenstein_2004} of known topology and branch lengths with the $n$ leaves being the observed sequences. The sequences at the other $n-1$ nodes are unobserved. Nodes are numbered backwards in time, starting from the observed leaves (numbered 1 to $n$) to the root $2n{-}1$. Let $\nu:[2n-2]\rightarrow\{n{+}1,\ldots,2n-1\}$ map nodes other than the root onto their parent node. It is assumed that we are given a Markov model on the tree describing the evolution of states over time on each branch of the tree and at each site; each site evolves independently given the branch lengths. Treating the sequence states at internal nodes as missing data we can write the full-data likelihood as:
\begin{equation}\label{eq:fullik}
p(x_{1:2n-1,1:m}|\theta) = \prod_{j=1}^m \mu_{\theta}(x_{(2n-1)j}) \prod_{i=1}^{2n-2} f_\theta(x_{ij}|x_{\nu(i)j})
\end{equation}
where $\theta\in\Theta\subseteq\mathbb{R}^d$ is an unknown parameter, $f_{\theta}$ a Markov transition probability,
and $\mu_{\theta}$ a probability distribution on the state space at a site. Note that $x_{n+1:2n-1,1:m}$ are unobserved, whereas $x_{1:n,1:m}$ are observed. Note also that $f_\theta(x_{ij}|x_{\nu(i)j})$ can depend on the (known) length of the branch connecting the node $i$ with node $\nu(i)$. For convenience in subsequent formulas we will write $f_\theta(x_{(2n-1)j}|x_{\nu(2n-1)j})$ in place of $\mu_{\theta}(x_{(2n-1)j})$, even though $\nu(2n{-}1)j$ is undefined.

The observed-data likelihood can be written as a sum over the missing data:
\begin{equation}\label{eq:obslik}
p(x_{1:n,1:m}|\theta) = \prod_{j=1}^m\left[\sum_{x_{n+1:2n-1,j}\in[p]^{n-1}} \prod_{i=1}^{2n-1} f_\theta(x_{ij}|x_{\nu(i)j})\right].
\end{equation}
Using belief propagation \cite{Pearl_1982} (also called the sum and products algorithm), the cost of computing \eqref{eq:obslik} is $\mathcal{O}(mp^2 n^2)$.

%which uses the tree structure to reduce the $\mathcal{O}(mp^n)$ cost of computing the marginals to $\mathcal{O}(mp^2 n^2)$.

Our model generalises (\ref{eq:fullik}) to allow $\theta$ to vary along the sequence at a set of change-points $1=s_0<s_1<\cdots<s_{k+1}=m$. Then the full-data likelihood for the change-point model is:
$$
p(x_{1:2n-1,1:m}|k,s_{1:k},\theta_{1:k+1}) = \prod_{j=1}^{k+1} \prod_{l=s_{j-1}}^{s_j - 1} \prod_{i=1}^{2n-1} f_{\theta_j}(x_{il}|x_{\nu(i)l}),
$$
and, as in (\ref{eq:obslik}), one can sum over $x_{n+1:2n-1,1:m}$ to obtain an observed-data likelihood:
$$
p(x_{1:n,1:m}|k,s_{1:k},\theta_{1: k+1}) = \prod_{j=1}^{k+1} \prod_{l=s_{j-1}}^{s_j - 1} p(x_{1:n,l}|\theta_j)
$$
where
$$
 p(x_{1:n,l}|\theta_j) = \sum_{x_{n+1:2n-1,l}\in[p]^{n-1}} \bigg\{\prod_{i=1}^{2n-1} f_{\theta_j}(x_{il}|x_{\nu(i)l})\bigg\}.
$$

\subsubsection{Bayesian Inference}\label{sec:bayesianinference}
For $ 0\leq k < m$, let 
$$
\mathsf{S}_k = \{s_{1:k}\in[m]: 1 < s_1<\cdots < s_k \leq m\}.
$$
Then we will define a posterior probability on the space
$$
\mathsf{E} := \bigcup_{k= 0}^{m-1}\Big( \{k\}\times\mathsf{S}_k\times\Theta^{k +1}\Big).
$$
Let $p(k,s_{1:k},\theta_{1:k+1})$ be any proper prior probability on $\mathsf{E}$. Our objective is then to consider the posterior
\begin{equation}\label{eq:posterior}
\pi(k,s_{1:k},\theta_{1: k+1}) \propto 
 p(x_{1:n,1:m}|k,s_{1:k},\theta_{1: k+1}) p(k,s_{1:k},\theta_{1: k+1})
\end{equation}
which can be computed pointwise up to a normalizing constant in $\mathcal{O}(mp^2 n^2)$ steps. We assume that we know how to calculate the priors $p(s_{1:k},\theta_{1: k}|k)$ and $p(k)$.

\subsubsection{Time Machine}\label{sec:timemachine}
One way to cut the cost of the $\mathcal{O}(mp^2 n^2)$ calculation of the likelihood, is to remove the top of the tree 
(a related idea is used in \cite{Jasra_2011} for the standard coalescent). Suppose we only consider the tree backward in time until the parent of node $2n{-}g$, $g\in\{2,\dots,n\}$. We propose the model:
\begin{align*}
 p^B &(x_{1:2n-g,1:m},x_{2n-g+1:2n-1,1:m}|k,s_{1:k},\theta_{1: k+1}) = \\
 &\prod_{j=1}^{k+1} \prod_{l=s_{j-1}}^{s_j - 1}\Big( \Big\{\prod_{i=1}^{2n-g} f_{\theta_j}(x_{il}|x_{\nu(i)l})\Big\}\eta_{\theta_j}\big(x_{ \mathcal{B}(2n-g+1:2n-1),l}\big)\Big),
\end{align*}
where $\mathcal{B}(2n-g+1:2n-1)$ denotes the nodes in the cut-off part of the tree $2n-g+1:2n-1$ that are parents to at least one of the nodes in $1:2n-g$, and $\eta_{\theta_j}(\cdot)$ is a joint probability distribution over sequences on these `boundary' nodes. Thus, the joint distribution of a number of the upper-most $g-1$ nodes, for the $j$ site, is replaced by the approximation $\eta_{\theta_j}$. Then one can perform inference from the relevant posterior
$$
\pi^{B}(k,s_{1:k},\theta_{1: k+1}) \propto 
 p^B(x_{1:n,1:m}|k,s_{1:k},\theta_{1:k +1}) p(k,s_{1:k},\theta_{1:k+1})
$$
using the PMMH method described below. The cost of computing the new likelihood is now $\mathcal{O}(mp^2 n(n-g))$.

%We will now consider the bias in using the time machine approximation.
%Let $\pi_{\theta_j}(x_{\nu(2n-g)l})$ denote the true marginal induced by the model, and suppose (see e.g.~\cite[Chapter 4]{delmoral}) that there exists an $\epsilon\in(0,1)$ such that, for every $\theta\in\Theta$, $x',x,y\in[p]$,
%$$
% f_{\theta}(x'|x) \geq \epsilon f_{\theta}(x'|y)
%$$
%with $ f_{\theta}(x'|x)>0$ for every $\theta,x,x'$. If this assumption is too strong, then, one can restate the assumption on the iterated kernel; this is not done purely for notational convenience.
%We let $\mathcal{B}_b(\mathsf{E})$ be the bounded and measurable functions on $\mathsf{E}$ and $\|\cdot\|$ denote the total variation distance.
%We have the following result, whose proof follows from standard results on Dobrushin coefficients (see again \cite[Chapter 4]{delmoral}):
%\begin{prop}
%There exists a $C<\infty$ such that for any $\zeta\in\mathcal{B}_b(\mathsf{E})$, $g\in[n]$
%$$
%\Big|\int_{\mathsf{E}}\zeta(k,s_{1:k},\theta_{1:k})[\pi(k,s_{1:k},\theta_{1:k}) - \pi^B(k,s_{1:k},\theta_{1:k})] d(k,s_{1:k},\theta_{1:k})\Big| \leq
%$$
%$$
%C \sup_{u\in\mathsf{E}}|\zeta(u)|(1-\epsilon)^{m(2n-g-1)} \int_{\mathsf{E}} \Big(\prod_{j=1}^{k+1} \prod_{l=s_{j-1}}^{s_j - 1}\|\pi_{\theta_j}-\eta_{\theta_j}\|\Big)p(k,s_{1:k}\theta_{1:k})d(k,s_{1:k},\theta_{1:k}).
%$$
%\end{prop}
%Essentially the bias falls as $g$ gets smaller and at a geometric rate. However, one cannot remove the bias unless $\eta_{\theta}=\pi_{\theta}$.

\subsection{Particle Marginal Metropolis-Hastings (PMMH)}\label{sec:Computationalmethods}
In order to sample from the trans-dimensional state-space of \eqref{eq:posterior} we first consider an SMC sampler which only samples on $\mathsf{S}_k\times\Theta^{k+1}$, for $k$ fixed. We then show how the SMC sampler can be embedded within a PMMH algorithm to target \eqref{eq:posterior}.
The SMC sampler will be necessary to ensure a good acceptance probability for trans-dimensional moves. Our approach has the advantage over alternative simulation techniques for model selection (see \cite{zhou}) that the model selection and parameter estimates are simultaneous, which helps to focus computational resources on the important model(s).

For $1\leq k < m$, and a user-specified $T\ge 1$, let $\{\xi_{t,k}\}_{0\leq t \leq T}$ be a sequence of probabilities
on $\mathsf{S}_k\times\Theta^{k+1}$, such that $\xi_{0,k}(s_{1:k},\theta_{1:k+1}) = p(s_{1:k},\theta_{1:k+1}|k)$ and 
$$
\xi_{T,k}(s_{1:k},\theta_{1:k+1}) \propto p(x_{1:n,1:m}|k,s_{1:k},\theta_{1:k+1}) p(s_{1:k}\theta_{1:k+1}|k).
$$
The remaining sequence of targets $\{\xi_{t,k}\}_{1\leq t \leq T-1}$ interpolate between the (conditional) posterior and the prior, e.g.\@ via the tempering procedure:
$$
\xi_{t,k}(s_{1:k},\theta_{1:k+1}) \propto p(x_{1:n,1:m}|k,s_{1:k},\theta_{1:k+1})^{\kappa_t} p(s_{1:k},\theta_{1:k+1}|k)
$$
with $0<\kappa_1<\cdots<\kappa_{T-1}<1$. The SMC sampler will propagate a collection of $N$ particles 
from the prior $\xi_{0,k}$ all the way to the posterior $\xi_{T,k}$ via the bridging densities $\xi_{t,k}$ by means 
of importance sampling, resampling and MCMC move steps. The tempering procedure aims at controlling 
the variability of the incremental importance weights, for instance providing robust estimates 
of the normalising constants $p(x_{1:n,1:m}|k)$ which is an important attribute for the overall algorithm. The sampler propagates the particles by using a sequence of MCMC kernels of invariant densities $\xi_{t,k}$ (which operate on a fixed dimensional space). All the details of the specific steps of the SMC sampler are given in the Supporting Information document. We write the probability of all the variables associated to the SMC sampler (which resamples $N>1$ `particles' at every time except at time $T$) as
$$
\Psi_{k,N}(a_{0:T-1}^{1:N},\phi_{0:T}^{1:N}(k)),
$$
where $a_{0:T-1}^{1:N} = (a_0^1,\dots,a_0^N,\dots,a_{T-1}^1,\dots,a_{T-1}^N)$ are the resampled indices and $\phi_{0:T}^{1:N}(k) = (\phi_0^1(k),\dots,\phi_0^N(k),\dots,\phi_{T}^1(k),\dots,\phi_{T}^N(k))$
with $\phi_t^i(k) = (s_{t,1:k}^i,\theta_{t,1:k+1}^i)$, $i\in[N],t\in\{0\}\cup[T]$, is the collections of the $N$ particles as propagated through the sequence $\xi_{t,k}$.

One can use this SMC sampler within a broader PMMH algorithm to sample from the true target of interest \eqref{eq:posterior}. The specific steps of PMMH are given in the Supporting Information document, but briefly, a single iteration of the algorithm is as follows. Given the current state of the Markov chain, one proposes to change $k$ with some proposal kernel
$q(k'|k)$. Conditional on this $k'$, we run an SMC sampler $\Psi_{k',N}(\cdot)$ and choose a particle $\phi_T^l(k')$, for some $1\le l\le N$, with probability proportional to a weight. Acceptance
of both the model index $k'$ and of the proposed change-point times and rates $\phi_T^l(k')$ happens with probability
$$
1\wedge \frac{p^N(x_{1:n,1:m}|k')p(k')}{p^N(x_{1:n,1:m}|k)p(k)}\times \frac{q(k|k')}{q(k'|k)},
$$
where $p^N(x_{1:n,1:m}|k)$ is the SMC (unbiased) estimate of $p(x_{1:n,1:m}|k)$, the normalizing constant of $\xi_{T,k}$. The Supporting Information document presents the formula used to calculate $p^N(x_{1:n,1:m}|k)$. Note that whilst there are a lot of user set parameters (namely, the temperatures
and tuning parameters for the MCMC kernels), their choice can be done 
adaptively to reduce user involvement
%via some preliminary runs for the different choices of the discrete variable $k$ say with a larger 
%number of particles compared to the one used in the final run 
(see
\cite{jasra}). In this article we tune the parameters by trial and error.

The advantages of our procedure is that it mitigates having to construct trans-dimensional proposals which need to mix well (see \cite{Karigiannis_2013} for another recent work that attempts to deal with this issue). We note, however, that the cost of each proposal will be $\mathcal{O}(NTmp^2 n^2)$, as $\xi_{t,k}$ must be obtained at each time step of the SMC sampler (see Supporting Information). In addition, note that tailored methods for change-point models (e.g.~\cite{Fearnhead_2007}) do not apply here as one does not have a convenient way to integrate the likelihood.

\subsection{Approximate Bayesian Computation (ABC)}\label{sec:abc}
ABC is another methodology that avoids exact computation of the likelihood, at the cost of a biased approximation of the posterior; see for instance \cite{marin} for a review. The method is based on accepting simulated data sets that are similar to the observed dataset, where `similar' is usually assessed using summary statistics sensitive to the parameter(s) of interest.

ABC can be unreliable as a tool for model selection. According to \cite{Marin_2013}, the best summary statistics to be used in ABC approximation to a Bayes factor are ancillary statistics with different mean values under two competing models. Otherwise, the summary statistic must have enough components to prohibit a parameter under a wrong model from generating summary statistics that are plausible under the true model. However, summary statistics satisfying the conditions of \cite{Marin_2013} for model choice in ABC is not easy (or even possible) to verify in our context.

In the numerical examples of Section \ref{sec:resultscomparison}, we consider two ABC algorithms which approximate the same ABC posterior. The first algorithm is a PMMH that replaces the SMC sampler of \cite{DelMoral_2006} with the SMC sampler of \cite[Section 3.3]{DelMoral_2012}; see Supporting Information for details. The second ABC algorithm is the ABC-SMC algorithm for model selection appearing on \cite[page 190]{Toni_2009}.
%At each time step and for each particle of this latter method, the data is simulated $M$ times and the proportion of simulated samples which are deemed sufficiently close to the true data is used to calculate the acceptance ratio of the sampler's MCMC kernel and also the weight of a particular particle at a given point in time.

%The second ABC algorithm is the ABC-SMC algorithm for model selection appearing on \cite[page 190]{Toni_2009}. In the context of our model, the scheme maintains separate populations of weighted samples of $(s_{1:k},\theta_{1:k})$ per value of $k$, for a total population of $N$ particles. At each of the algorithm's $T$ iterations, we build a new population of $N$ particles through resampling and importance sampling techniques, with the intention of obtaining a more accurate population as $T$ increases. ABC is used to accept or reject newly sampled values of $(s_{1:k},\theta_{1:k})$, so that we only assign weights to parameter values which are sampled from a biased posterior. One potential issue with this ABC-SMC algorithm is that it requires $\mathcal{O}(N^2)$ operations (all of the other algorithms discussed thus far are $\mathcal{O}(N)$ in complexity). Thus, there will likely be a limit on the number of particles that one could realistically use.

\section{Results}\label{sec:results}

\subsection{Comparison of Computational Methods on Simulated Data}\label{sec:resultscomparison}
We compared three algorithms on their performance in Bayesian model selection for four simulated DNA datasets. Within each dataset, the DNA sequences shared a common ancestral binary tree with known topology, unknown sequence states at ancestral nodes, and unknown substitution rates and branch lengths. The first algorithm was our proposed PMMH algorithm outlined in Section \ref{sec:Computationalmethods}, and we employed three versions of the time machine.  Using the notation of Section~\ref{sec:timemachine}, these used $g=1$ (so in effect the time machine was not implemented at all), $g=4$ and $g=8$. We also used two ABC algorithms described in Section \ref{sec:abc}.  The PMMH algorithms were not run until they converged fully, but were compared on the basis of results achieved after six hours of computation.  Other implementation details of the algorithms may be found in the Supporting Information document.

\subsubsection{Base dataset}
The base dataset consists of $n=8$ simulated DNA sequences ($p=4$ types of nucleotide), each of $m=50$ sites.  The sequences evolved according to a binary tree under a Jukes-Cantor model of DNA evolution with one substitution rate up to site $s_1=25$, and a second rate beyond this single change-point (so $k=1$, but for inference we assumed only $k\in\{0,1\}$). In the standard Newick notation, the structure of the tree was:

\verb+(((Taxon0:1.0,Taxon1:1.0):1.0,(Taxon2:1.0,Taxon3:1.0):1.0):1.0,+

\verb+((Taxon4:1.0,Taxon5:1.0):1.0,(Taxon6:1.0,Taxon7:1.0):1.0):1.0):1.0+

We ran the three algorithms to infer $k$, location of the change-point, $s_1$ given $k{=}1$, and the substitution rate(s) $\theta_{1:k+1}$. The prior on $k$ was uniform on $\{0,1\}$; the prior on $s_1$ was $1/m{-}1$ (change-points accur immediately before a site so cannot occur at site 1; finally, all substitution rates had a gamma prior, with shape $=2$ and scale $=0.4$, and so expected value of $0.8$ mutations per generation per site.

The results in the top quadrants of Tables \ref{tab:modelchoiceex1} and \ref{tab:CIex1} in Appendix \ref{sec:tables} show that our time-machine PMMH algorithm with $g=4$ outperformed all other algorithms. It sampled from the true model (i.e., $k=1$) much more frequently than the incorrect $k=0$ model (Table \ref{tab:modelchoiceex1}). In comparison $g=8$ performed poorly, as expected since $n=8$ for this dataset so $g=8$ implies removing all internal nodes and assuming independent evolution of each sequence. The ABC algorithms did not perform well. The PMMH-ABC algorithm sampled from the two models almost evenly, while the ABC-SMC algorithm had a low effective sample sizes (\cite{Kong_1994},\cite{Liu_1996}) and actually preferred the wrong model.

In Table \ref{tab:CIex1}, we give $95$\% confidence intervals of estimates of $s_1$ and of the rates given $k=1$. The time-machine PMMH algorithms again provide the best inferences and were able to find the change-point. The PMMH-ABC was more accurate for the substitution rates but less precise. The ABC-SMC algorithm gave unusable output.

We do not present the output for the $g=1$ version of the time machine because it performed very poorly. Without removing any nodes from the top of the tree, the variability of ${p^N(x_{1:n,1:m}|k')}/{p^N(x_{1:n,1:m}|k)}$ in the acceptance probability of the PMMH was very high when $k\neq k'$ (Figure \ref{fig:ML} in Section \ref{sec:figures}). Thus, the algorithm accepted jumps between models only rarely and the output was very ``sticky''. This phenomenon illustrates that the time machine is a cost saving technique by two measures. First, it reduces the computational complexity of the algorithm. Second, it aids in mixing and facilitates jumping between models.

\subsubsection{Further Tests}
We repeated the above experiment for three more datasets that differed only slightly from the base dataset. We found the results to be similar across the datasets (see Tables \ref{tab:modelchoiceex1} and \ref{tab:CIex1} in Section \ref{sec:tables}). Collectively, these results suggest that when doing Bayesian model selection under these scenarios, ABC approximations should be avoided and instead our PMMH method used instead, with the time machine but removing as few nodes as computational considerations permit.

\subsection{Application to a Real Dataset}\label{sec:resultsrealdata}
%The results of Section \ref{sec:resultscomparison} suggest that the PMMH algorithm which employs the time machine is the relatively superior computational method when one is performing inference on these binary tree models. Thus, we further tested that algorithm on actual data. We found that the algorithm was able to converge and give precise output, as the following paragraphs will explain.

Using the publicly available database of \cite{Weiss_2013}, we assembled a dataset consisting of $n=6$ ACT1 gene DNA sequences ($m=540$ sites). We assumed the tree structure given in Figure \ref{fig:AppliedInputTree} of Section \ref{sec:figures}, and a Jukes-Cantor model of DNA evolution. %(see \cite[Chapter 13]{Felsenstein_2004}). 
We implemented our time-machine PMMH algorithm to infer $k$, $s_{1:k}$, and $\theta_{1:k+1}$ for cut-off parameter $g=4$ (see Supporting Information for further details). The prior on $k$ was a discrete uniform distribution on $\{0,1,2\}$, the prior on $s_{1:k}$ was uniform on $k$-subsets of $[m{-}1]$, and each substitution rate had a gamma prior with shape $=1$, scale $=0.3$.

We ran the algorithm for 10,589 iterations ($23$ days) on a Linux workstation that used twelve Intel Xeon E5-1650 3.20 GHz CPUs. We monitored convergence via autocorrelation and trace plots (Figure \ref{fig:kACFTrace}). We also monitored convergence of each model individually using the diagnostic in \cite{Geweke_1992}; that is, we obtained a Z-score for each model parameter per each value of $k$ to get a sense of the algorithm's ability to fully explore the state space of each model (only some values are reported below).  Figure \ref{fig:kACFTrace} suggests good exploration of the state space of $k$, resulting in an estimated distribution: $0.17$ ($k=0$), $0.47$ ($k=1$), and $0.36$ ($k=2$).

From the 4,946 samples with $k=1$, we estimated a 95\% highest posterior density interval of (194,199) (see also the histogram in Figure \ref{fig:ksplots}). The Z-score for $k$ was $-0.60$, suggesting that we were still some way off convergence (values close to $0$ imply convergence).  For the rates $\theta_1$ (before the change-point) and $\theta_2$ (after the change-point), the  Z-scores of $0.11$ and $0.23$, respectively, give stronger evidence for convergence (estimated densities of these parameters are shown in Figure \ref{fig:rateplot}).

\section{Discussion}\label{sec:disc}
We considered sequence data that originates from a rooted binary tree \cite[Chapter 1]{Felsenstein_2004} of known topology and branch lengths but unknown sequence states at internal nodes, and we the substitution rates in the DNA evolution model allowed to have change-points. We detailed Bayesian parameter inference from such a model with an unknown number of change-points, implying a trans-dimensional posterior density. Computational inference from this model is challenging, and we introduced two novel contributions to facilitate sampling.

Firstly, based on the time machine principle of \cite{Jasra_2011}, we showed how the top-most nodes of the binary tree can be replaced with a probability distribution of the sequence evolution model to reduce the cost of computing the likelihood linearly in $n$ (the number of sequences). This approach introduces a bias, but this is was found in practice to have a small effect on inferences.

Secondly, we developed a particle marginal Metropolis-Hastings (PMMH) algorithm  (Section \ref{sec:Computationalmethods}) which mitigates having to construct trans-dimensional proposals that need to mix well. We first developed a sequential Monte Carlo (SMC) sampler which only samples on a fixed-dimensional subspace of the full trans-dimensional state-space. We then showed how that SMC sampler can be embedded within the PMMH algorithm to target the full posterior. By employing the time machine within this PMMH, we attained an algorithm that could run with a reduced computational cost and easily jump between models with different numbers of change-points.

We successfully implemented our PMMH to perform inference from the model in a reliable fashion for small to moderately sized datasets. We empirically demonstrated that our PMMH can outperform approximate 
Bayesian computation (ABC) techniques \cite{Tavare_1997} in terms of precision and accuracy, and we showed that our algorithm can successfully be used to carry out reliable inference on real data. The success of our PMMH algorithm is largely due to the time machine, which, as we witnessed in Section \ref{sec:resultscomparison}, reduces the variance of the acceptance probability and enables the algorithm to jump easily between models. However, based on the output of Section \ref{sec:resultscomparison}, it seems that bias introduced by the time machine reduces the accuracy of the inferred substitution rates.

In a future work, one might want to extend the methodology to allow for unknown tree topologies, similar to \cite{suchard}. Also, a future work could attempt to use a more appropriate distribution to approximate the distribution at the top of the tree. We attempted to find approximations in the point processes and coalescent literature, but we were unable to find a better approximation than that which we employed here. From the computational point of view, it will certainly be important to further speed up the algorithm and great savings could be made by parallelising calculations within the SMC particle method and carefully investigating adaptive procedures for fine-tuning the temperatures and the MCMC kernels. All such efforts could have a big effect on reducing the variance of the estimate of $p(x_{1:n,1:m}|k)$, thus further improving the mixing of PMMH even with fewer removed nodes. Also, there could then be great scope to apply the method for larger number of potential change points compared to 
the relatively small one we have tried here.

\section*{Acknowledgements}
This research was funded by the EPSRC grant ``Advanced Stochastic Computation for Inference from Tree, Graph and Network Models'' (Ref: EP/K01501X/1). AJ was additionally supported by Singapore MOE grant R-155-000-119-133 and is also affiliated with the risk management institute at the National University of Singapore.

\appendix
\section{Figures}\label{sec:figures}
\begin{figure}[H]
\begin{center}
\textbf{Real Data Case: Phylogeny of a subset of the Saccharomycotina subphylum} \\ [1.2ex]
\scalebox{0.75}{\includegraphics[trim = 0mm 0mm 0mm 0mm, clip]{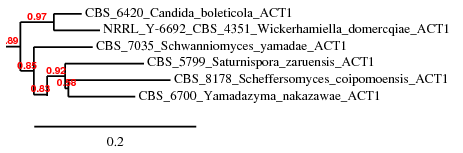}}
\caption{This tree is provided by \cite{Weiss_2013}. The labels on the leaves at far right are in the format STRAIN\textunderscore taxa\textunderscore GENE. For example, ``CBS\textunderscore 6420\textunderscore Candida\textunderscore boleticola\textunderscore ACT1'' is the label for the ACT1 gene of \textit{Candida boleticola}, strain CBS 6420. The edge at the bottom of the figure with the number $0.2$ shows the branch length that represents an amount of genetic change equal to $0.2$ nucleotide substitutions per site. The red numbers represent a measure of evidence for the node, with values closer to $1.0$ meaning that there is stronger evidence for the genes to the right of the node clustering together; in this case, those values are computed using a bootstrapping method.}
\label{fig:AppliedInputTree}
\end{center}
\end{figure}

\begin{figure}[H]
\begin{center}
\textbf{Sim. Data base dataset: Variability of $\log\left[p^N(x_{1:n,1:m}|k)\right]$ for $g=1$ versus $g=4$} \\
\scalebox{0.5}{\includegraphics[trim = 10mm 10mm 5mm 100mm, clip]{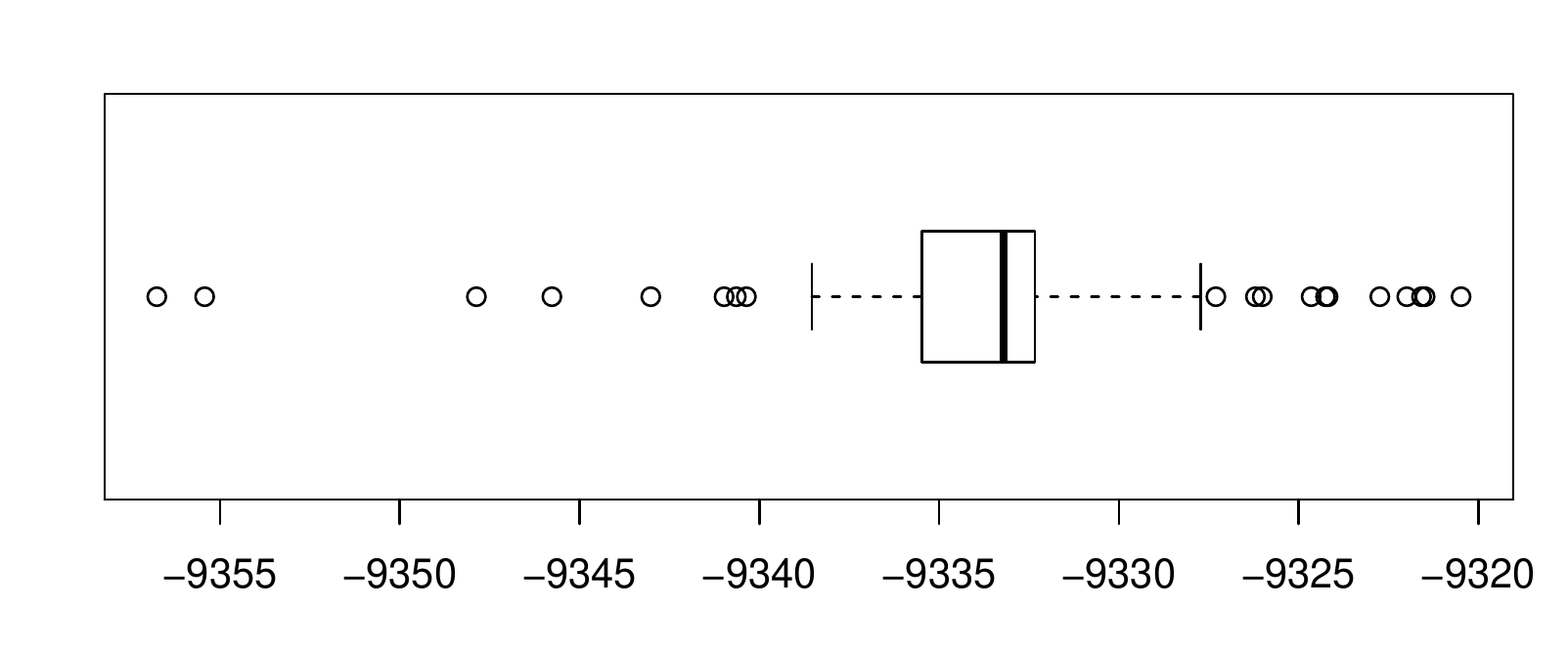}}\\
\scalebox{0.5}{\includegraphics[trim = 10mm 100mm 5mm 20mm, clip]{ML.pdf}}
\vspace{-0.3cm}
\caption{The top box plot illustrates the variability of $\log\left[p^N(x_{1:n,1:m}|k)\right]$ for the $g=1$ version of our algorithm, and the bottom plot gives the same for the $g=4$ version of our algorithm; $\log\left[p^N(x_{1:n,1:m}|k)\right]$ runs along the horizontal axis of each plot. One can see that the variability in the bottom figure is much less than that of the top figure.}
\label{fig:ML}
\end{center}
\end{figure}

\begin{figure}[H]
\begin{center}
\textbf{Real Data Case: Autocorrelation and trace plot of sampled $k$} \\ [1.2ex]
\scalebox{0.5}{\includegraphics[trim = 0mm 0mm 0mm 20mm, clip]{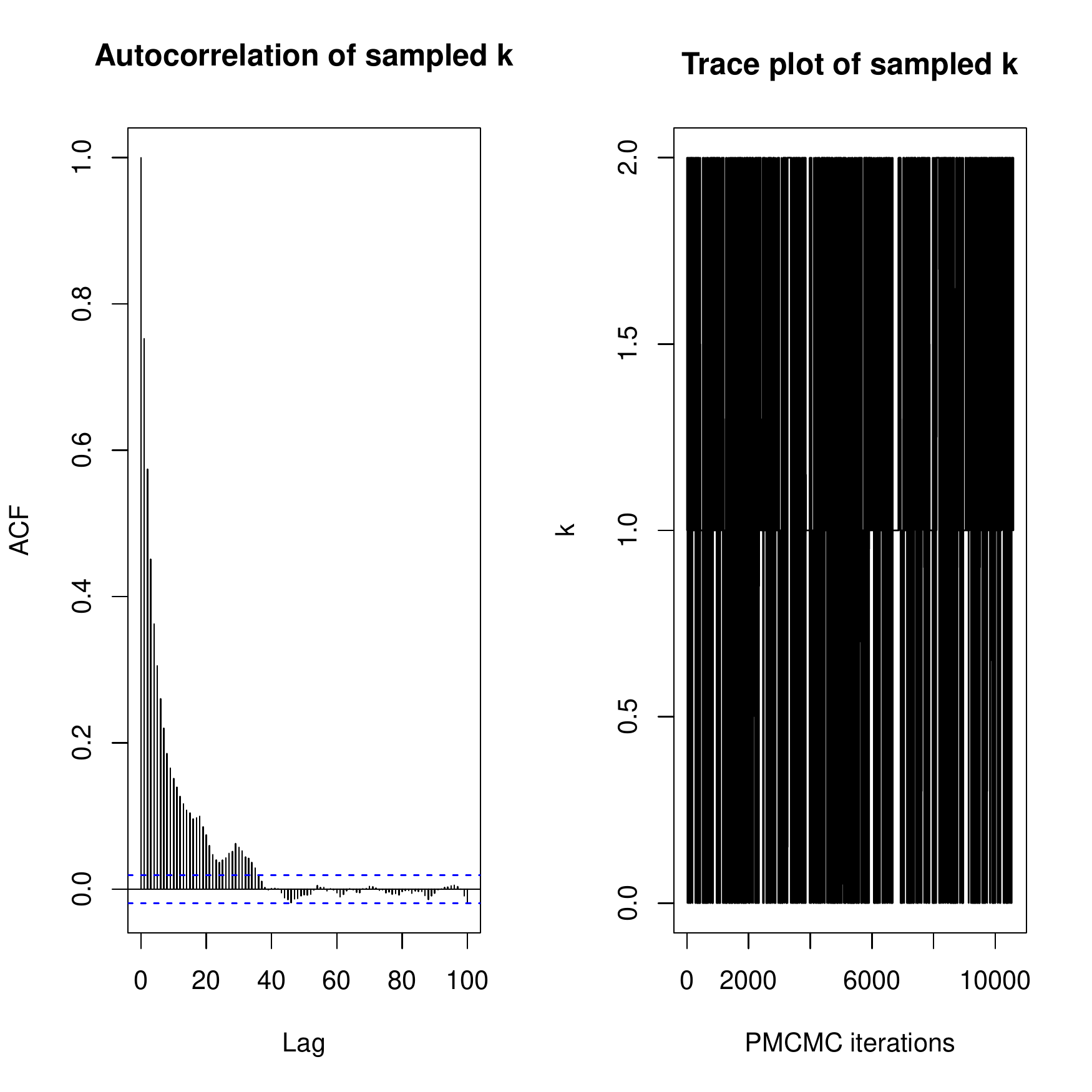}}
\vspace{-0.3cm}
\caption{We monitored the convergence of the PMMH implemented in Section \ref{sec:resultsrealdata} via autocorrelation and trace plots. The plots illustrate a non-sticky algorithm and low autocorrelation.}
\label{fig:kACFTrace}
\end{center}
\end{figure}

\begin{figure}[H]
\begin{center}
\textbf{Real Data Case: Histogram of sampled $s_1$ given $k=1$} \\
\scalebox{0.5}{\includegraphics[trim = 0mm 10mm 0mm 14mm, clip]{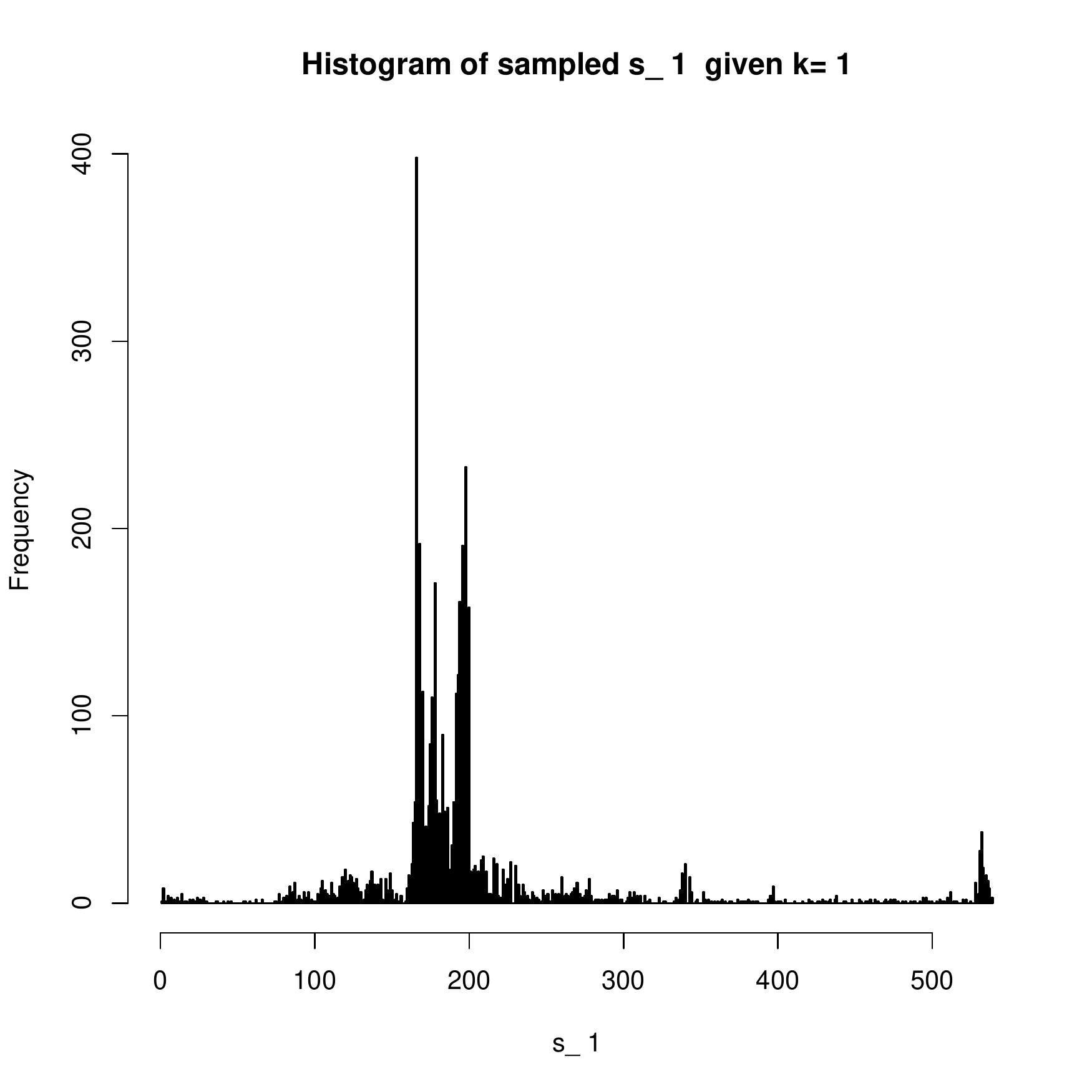}}
\vspace{-0.3cm}
\caption{The histogram strongly indicates the position of the single change-point in the numerical example of Section \ref{sec:resultsrealdata}.}
\label{fig:ksplots}
\end{center}
\end{figure}

\begin{figure}[H]
\begin{center}
\textbf{Real Data Case: Kernel density plots of sampled substitution rates given $k=1$} \\
\scalebox{0.5}{\includegraphics[trim = 0mm 100mm 0mm 10mm, clip]{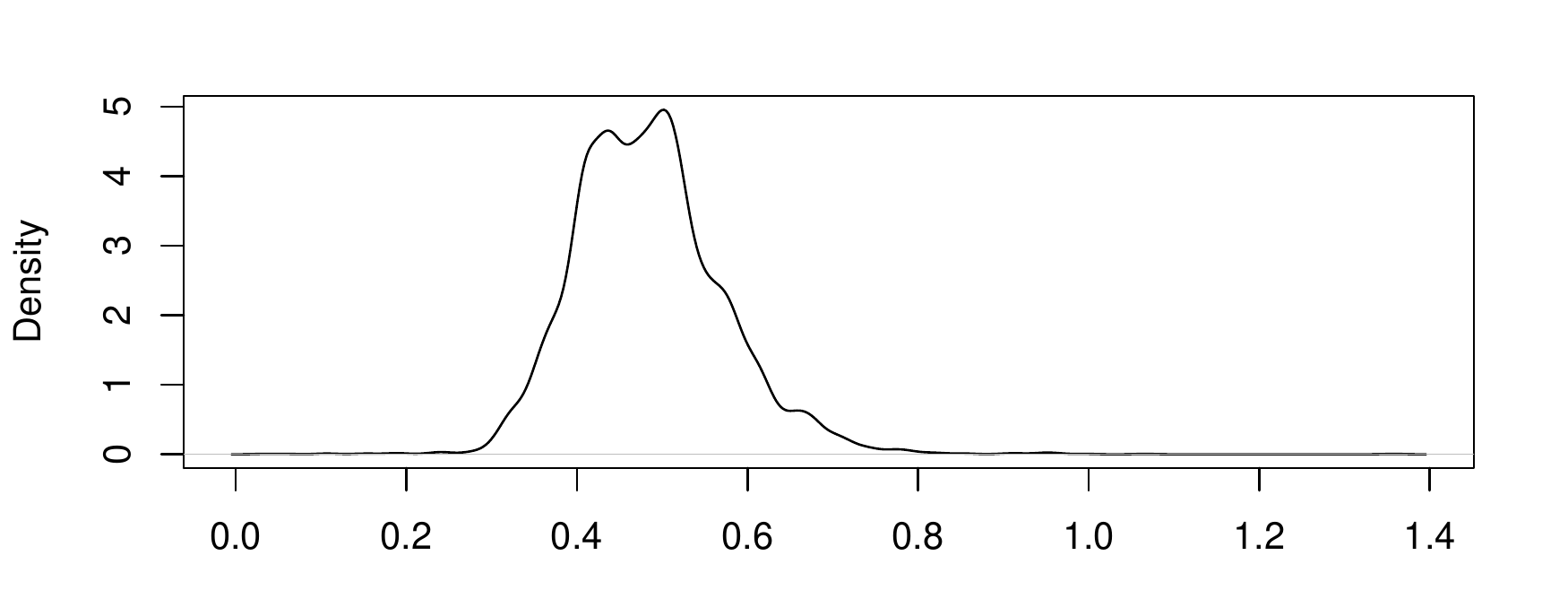}} \\
\scalebox{0.5}{\includegraphics[trim = 0mm 0mm 0mm 100mm, clip]{rateplot.pdf}}
\vspace{-0.3cm}
\caption{At top, we have the kernel density plot of $\theta_1$, which is obtained using all samples from the PMMH implemented in Section \ref{sec:resultsrealdata} where $k=1$. Values for $\theta_1$ run along the horizontal axis. We have the same plot for $\theta_2$ at bottom.}
\label{fig:rateplot}
\end{center}
\end{figure}

\section{Tables}\label{sec:tables}
\begin{sidewaystable}[!h]
 \caption{Algorithm diagnostics and model choice}\label{tab:modelchoiceex1}
 \centering
 \begin{tabular}{l l c c cc c ccc c ccc}
 \hline\hline
 & & & Accp. & \multicolumn{2}{c}{Autocorrelation of $k$} & & \multicolumn{3}{c}{ESS per model} & & \multicolumn{3}{c}{$\mathbb{P}(k \mid \text{data})$} \\
 \cline{5-6} \cline{8-10} \cline{12-14}
 Example & Algorithm & Samples & ratio & @25 & @100 & & $k=0$ & $k=1$ & $k=2$ & & $k=0$ & $k=1$ & $k=2$ \\
 \hline
 base dataset & Time machine, $g=4$ & 4255 & 0.214 & 0.029 & 0.035 & & -- & -- & -- & & 0.261 & 0.739 & -- \\
 ($k=1$, & Time machine, $g=8$ & 7237 & 0.325 & -0.001 & -0.006 & & -- & -- & -- & & 0.477 & 0.523 & -- \\
$n=8$, red{$m=50$}) & PMMH with ABC & 1070 & 0.305 & -0.012 & 0.025 & & -- & -- & -- & & 0.482 & 0.518 & -- \\
 & ABC-SMC & 5000 & -- & -- & -- & & 6 & 7 & -- & & 0.538 & 0.462 & -- \\ [1ex]
 \hline
 Two change- & Time machine, $g=4$ & 4213 & 0.121 & 0.184 & 0.036 & & -- & -- & -- & & -- & 0.159 & 0.841 \\
 points & Time machine, $g=8$ & 7198 & 0.287 & -0.004 & -0.001 & & -- & -- & -- & & -- & 0.472 & 0.528 \\
 ($k=2$, & PMMH with ABC & 1021 & 0.394 & 0.010 & 0.011 & & -- & -- & -- & & -- & 0.562 & 0.438 \\
$n=8$, $m=50$) & ABC-SMC & 5000 & -- & -- & -- & & -- & 1 & 1 & & -- & 0.477 & 0.523 \\ [1ex]
 \hline
 Subtle change- & Time machine, $g=4$ & 4264 & 0.108 & 0.017 & 0.027 & & -- & -- & -- & & 0.079 & 0.921 & -- \\
 point & Time machine, $g=8$ & 7229 & 0.306 & 0.005 & 0.014 & & -- & -- & -- & & 0.493 & 0.507 & -- \\
 ($k=1$, & PMMH with ABC & 1058 & 0.329 & 0.008 & 0.046 & & -- & -- & -- & & 0.486 & 0.514 & -- \\
$n=8$, $m=50$) & ABC-SMC & 5000 & -- & -- & -- & & 2 & 1 & -- & & 0.624 & 0.376 & -- \\ [1ex]
 \hline
 More sites & Time machine, $g=4$ & 530 & 0.106 & -0.015 & -0.037 & & -- & -- & -- & & 0.075 & 0.925 & -- \\
 ($k=1$, & Time machine, $g=8$ & 904 & 0.293 & -0.023 & -0.080 & & -- & -- & -- & & 0.461 & 0.539 & -- \\
$n=8$, $m=80$) & PMMH with ABC & 132 & 0.530 & 0.003 & 0.013 & & -- & -- & -- & & 0.469 & 0.531 & -- \\
 & ABC-SMC & 5000 & -- & -- & -- & & 22 & 7 & -- & & 0.508 & 0.492 & -- \\ [1ex]
 \hline
 \end{tabular}
 \\
 \begin{flushleft}
 All algorithms ran for six hours on a Linux workstation that used five Intel Core i5-2600 CPUs, each at 3.40 GHz (see important note regarding ABC-SMC in the Supporting Information document).
 The column labelled ``Accp. ratio'' gives the frequency at which a new sample is accepted in the particle marginal Metropolis-Hastings (PMMH) algorithms \cite{Andrieu_2010}.
 The columns labelled ``Autocorrelation of $k$'' give the autocorrelation at time delays of $25$ and $100$ (when available).
 ``ESS'' is the abbreviation for the effective sample size (\cite{Kong_1994},\cite{Liu_1996}) of the ABC-SMC for model selection of \cite{Toni_2009}. Note that for the ``two change-point'' case, the prior on $k$ was a discrete uniform distribution on $\{1,2\}$.
 %For each algorithm's estimate of the posterior of $k$, the greatest value of $\mathbb{P}(k \mid \text{data})$ is written in bold font.
 \end{flushleft}
\end{sidewaystable}

\begin{sidewaystable}
 \caption{Inference for true model: $95$\% confidence intervals}\label{tab:CIex1}
 \centering
 \begin{tabular}{l l c c c c c c}
 \hline\hline
 Example & Algorithm & Samples & {$s_1$} & {$s_2$} & {CI of $\mu_1$} & {CI of $\mu_2$} & {CI of $\mu_3$} \\
 \hline
 base dataset & Time machine, $g=4$ & 3144 & (22,23) & -- & (0.484,0.494) & (0.663,0.671) & -- \\
 ($s_1=25$, & Time machine, $g=8$ & 3784 & (25,26) & -- & (0.322,0.327) & (0.432,0.438) & -- \\
 $\mu_1=0.75$, $\mu_2=0.85$) & PMMH with ABC & 554 & (24,26) & -- & (0.709,0.811) & (0.702,0.794) & -- \\
 & ABC-SMC & 2310 & (1,1) & -- & (1.069,1.093) & (0.454,0.488) & -- \\ [1ex]
 \hline
 Two change-points & Time machine, $g=4$ & 3543 & (10,11) & (43,45) & (0.340,0.343) & (0.590,0.599) & (0.239,0.243) \\
 ($s_1=15$, $s_2=35$, & Time machine, $g=8$ & 3801 & (17,18) & (32,33) & (0.290,0.294) & (0.414,0.428) & (0.285,0.290) \\
 $\mu_1=0.75$, $\mu_2=0.85$, & PMMH with ABC & 447 & (15,22) & (34,42) & (0.712,0.807) & (0.698,0.795) & (0.708,0.799) \\
 $\mu_3=0.75$) & ABC-SMC & 2615 & (1,1) & (1,2) & (1e-7,1e-7) & (1e-7,1e-7) & (1e-7,1e-7) \\ [1ex]
 \hline
 Subtle change-point & Time machine, $g=4$ & 3927 & (21,22) & -- & (0.435,0.441) & (0.343,0.353) & -- \\
 ($s_1=25$, & Time machine, $g=8$ & 3665 & (22,24) & -- & (0.297,0.304) & (0.319,0.326) & -- \\
 $\mu_1=0.75$, $\mu_2=0.8$) & PMMH with ABC & 543 & (22,25) & -- & (0.678,0.774) & (0.690,0.788) & -- \\
 & ABC-SMC & 1880 & (2,2) & -- & (1.278,1.287) & (0.449,0.463) & -- \\ [1ex]
 \hline
 More sites & Time machine, $g=4$ & 490 & (41,46) & -- & (0.432,0.449) & (0.393,0.418) & -- \\
 ($s_1=40$, & Time machine, $g=8$ & 487 & (39,43) & -- & (0.426,0.439) & (0.406,0.425) & -- \\
 $\mu_1=0.75$, $\mu_2=0.85$) & PMMH with ABC & 70 & (34,45) & -- & (0.682,0.991) & (0.669,0.906) & -- \\
 & ABC-SMC & 2460 & (1,1) & -- & (1.289,1.345) & (0.505,0.531) & -- \\ [1ex]
 \hline
 \end{tabular}
 \\
 \begin{flushleft}
 The leftmost column contains the true parameter values for each example.
 We also record the number of samples on which each inference is based (i.e., we record how many samples from the true model each algorithm obtained).
 \end{flushleft}
\end{sidewaystable}

\clearpage

\section{Supporting Information: Algorithm Summaries}
\subsection{Sequential Monte Carlo (SMC) sampler}\label{sec:smcsamplersteps}
Consider the sequence $\{\xi_{t,k}\}_{0\leq t \leq T}$ introduced in Section 3.2 of the main paper, which is a sequence of probabilities known up-to a multiplicative constant. SMC samplers of \cite{DelMoral_2006} are designed for sampling from such sequences of distributions. The specific SMC sampler that we employ in this work simulates a collection of $N$ samples (or, `particles') in parallel and sequentially in time using a) a sequence of MCMC kernels of invariant densities $\xi_{t,k}$ and b) a resampling technique. The algorithm is summarized here as Algorithm \ref{alg:SMCsampler}, with $\phi_t^i(k) = (s_{t,1:k}^i,\theta_{t,1:k+1}^i)$, $i\in[N],t\in\{0\}\cup[T]$ and $T>0$ (note that each $\xi_{t,k}$ will be calculated via belief propagation \cite{Pearl_1982}). The $N$ outputted samples at time step $T$ provide an approximation of the target $\xi_{T,k}$. According to \cite[Section 3.2.1]{DelMoral_2006}, the unnormalized weights 
can be used to obtain an unbiased estimate of the normalizing constant of $\xi_{T,k}$:
\begin{equation}\label{eq:normalizingconstantest}
 p^N(x_{1:n,1:m}|k) = \prod_{t=0}^T \bigg[ \frac{1}{N} \sum_{i=1}^N W_t^i \bigg].
\end{equation}

\begin{algorithm}[H]
\begin{itemize}
\item{ Step 1: For $i\in[N]$, sample $\phi_0^i(k) \sim \xi_{0,k}(\cdot)$ and set the unnormalized weight: ${W}_0^i = 1$. Set $t=1$. }
\item{ Step 2: For $i\in[N]$, sample $a_{t-1}^i\in[N]$ from a discrete distribution on $[N]$ with $j^\text{th}$ probability
$w_{t-1}^j \propto W_{t-1}^j$. The sample $\{ a_{t-1}^{1:N} \}$ are the indices of the resampled particles at time step $(t-1)$. For each $i$, set the normalized weight ${w}_{t-1}^i=N^{-1}$. }
\item{ Step 3: For $i\in[N]$, sample $\phi_t^i(k) \mid \phi_{t-1}^{a_{t-1}^i}(k) \sim K_t\left( \cdot \mid \phi_{t-1}^{a_{t-1}^i}(k) \right)$, where $K_t$ is an MCMC kernel of invariant density $\xi_{t,k}$. Compute the unnormalized weight according to \cite[Equation 31]{DelMoral_2006}:
\begin{equation*}
{W}_t^i = \frac{\xi_{t,k}(\phi_{t-1}^{a_{t-1}^i}(k))}{\xi_{t-1,k}(\phi_{t-1}^{a_{t-1}^i}(k))}.
\end{equation*}
If $t=T$, stop. Otherwise, set $t=t+1$ and return to the start of Step 2.}
\end{itemize}
\caption{\label{alg:SMCsampler} Sequential Monte Carlo (SMC) sampler}
\end{algorithm}

\subsection{Particle Marginal Metropolis-Hastings (PMMH)}\label{sec:pmmhsteps}
Recall the target (3) of Section 3.1.1 of the main paper: $\pi(k,s_{1:k},\theta_{1:k+1})$. Particle Markov chain Monte Carlo (PMCMC) algorithms \cite{Andrieu_2010} consider an `extended target' that yields the true target of interest -- in this case, $\pi(k,s_{1:k},\theta_{1:k+1})$ -- as a marginal. The extended target is constructed in such a way that an SMC algorithm (e.g., Algorithm \ref{alg:SMCsampler}) can be used to sample some of its variables. In the context of this work, we will follow \cite{Andrieu_2010} and write an extended target as
\begin{align}\label{eq:extendedtarget}
 \pi^N &\left( l, k, a_{0:T-1}^{1:N},\phi_{0:T}^{1:N}(k) \right) = \\
 &\frac{\pi\left(k,\phi_{0:T}^{l}(k)\right)}{N^T} \cdot
 \frac{\Psi_{k,N}(a_{0:T-1}^{1:N},\phi_{0:T}^{1:N}(k))}{\xi_{0,k}(\phi_0^l(k)) \left(\prod_{t=1}^{T} {w}_{t-1}^{a_{t-1}^l}K_t\left(\phi_{t}^l(k)\mid \phi_{t-1}^{a_{t-1}^l}(k) \right)\right)}, \nonumber
\end{align}
where $\Psi_{k,N}$ is the probability of all the variables associated to the SMC sampler. A PMMH algorithm \cite{Andrieu_2010} is a type of PMCMC algorithm that can sample from \eqref{eq:extendedtarget}, and we present the details of the procedure as Algorithm \ref{alg:PMMH}.

\begin{algorithm}[H]
\begin{itemize}
 \item{ Step 0: Set $r=0$. Sample $k^{(r)}\sim p(k)$. All remaining random variables can be sampled from their full conditionals defined by the target \eqref{eq:extendedtarget}:
 
 - Sample $\phi_{0:T}^{(r),1:N}(k^{(r)}), a_{0:T-1}^{(r),1:N} \sim \Psi_{k^{(r)},N}(\cdot)$ via Algorithm \ref{alg:SMCsampler}.

 - Choose a particle index $l^{(r)} \propto W_T^{r,l^{(r)}}$.
 
 Finally, calculate $p^N(x_{1:n,1:m}\mid k^{(r)})$ via \eqref{eq:normalizingconstantest}.}
 
 \item{ Step 1: Set $r=r+1$. Sample $k^* \sim q\left( \cdot \mid k^{(r-1)} \right)$. All remaining random variables can be sampled from their full conditionals defined by the target \eqref{eq:extendedtarget}:
 
 - Sample $\phi_{0:T}^{*,1:N}(k^*), a_{0:T-1}^{*,1:N} \sim \Psi_{k^*,N}(\cdot)$ via Algorithm \ref{alg:SMCsampler}.

 - Choose a particle index $l^* \propto W_T^{*,l^*}$.

 Finally, calculate $p^N(x_{1:n,1:m}\mid k^*)$ via \eqref{eq:normalizingconstantest}.}

 \item{ Step 2: With acceptance probability
 \begin{align*}
 1\wedge \frac{p^N(x_{1:n,1:m}|k^*)p(k^*)}{p^N(x_{1:n,1:m}|k^{(r-1)})p(k^{(r-1)})}\times \frac{q(k^{(r-1)}|k^*)}{q(k^*|k^{(r-1)})},
 \end{align*}
 set $\Big( l^{(r)}, \,k^{(r)},\, \phi_{0:T}^{(r),1:N}(k^{(r)}), \,a_{0:T-1}^{(r),1:N} \Bigr) =\Big( l^*,\, k^*,\, \phi_{0:T}^{*,1:N}(k^*),\, a_{0:T-1}^{*,1:N}\Big)$. Otherwise, set 
$\Big( l^{(r)},\,k^{(r)},\,\phi_{0:T}^{(r),1:N}(k^{(r)}),\, a_{0:T-1}^{(r),1:N}\Big) = 
\Big(k^{(r-1)},\,\phi_{0:T}^{{(r-1)},1:N}(k^{(r-1)}),\,a_{0:T-1}^{{(r-1)},1:N}\Big)$.
 
 Return to the beginning of Step 1.}
\end{itemize}
\caption{\label{alg:PMMH}Particle marginal Metropolis-Hastings (PMMH)}
\end{algorithm}

\section{Supporting Information: Implementation Details for Section 4.1 }\label{sec:ImplementationDetailsEx1}
\subsection{Particle marginal Metropolis-Hastings (PMMH)}\label{sec:ImplementationDetailsEx1pmmh}
The PMMH algorithms \cite{Andrieu_2010} which employed the time machine all used $N=20$ particles, and the sequential Monte Carlo (SMC) sampler \cite{DelMoral_2006} within these algorithms always ran for $T=10$ time steps (for the ``More sites'' example, we set $T=50$). The PMMH employing the ABC \cite{Tavare_1997} algorithm of \cite{DelMoral_2012} also used $N=20$, but $M=20$ copies of the data were simulated for each particle in order to obtain the best results. Thus, this latter PMMH actually completed fewer iterations than the former PMMH within six hours.

In the numerical examples considered in this paper, we used the following proposals within our PMMH algorithms; note that these are only suggested proposals and the methodology is still valid with other choices. The number of change-points was propagated from PMMH iteration $(i-1)$ to iteration $i$ using a discrete uniform distribution which was centered on $k^{(i-1)}$ and had an odd integer width of $W>1$. In the event that
\begin{equation*}
 k^{(i-1)}-(W-1)/2<A \quad \text{or} \quad k^{(i-1)}+(W-1)/2>\Omega
\end{equation*}
(with $A$ and $\Omega$ being the lower and upper limits of possible values for $k$, respectively), then $k^{(i)}$ was sampled from a discrete uniform distribution with $W$ shortened as needed (i.e., any potential values for $k^{(i)}$ that would be out of bounds were removed from the support).

The SMC samplers within the PMMH propagated the change-points $s_{1:k}$ and the components of $\theta_{1:k+1}$ (i.e., the unknown substitution rates) via a Metropolis-Hastings kernel, and the unnormalised SMC sampler weights were calculated according to \cite[Equation 31]{DelMoral_2006}. Within the kernel, each individual change-point was propagated via a discrete uniform random walk similar to that which propagated $k$. To avoid selecting the same site twice at SMC iteration $i$, the sampling consisted of two steps:
\begin{enumerate}
 \item{ Sample $s_{t,1}$ conditional on $s_{t-1,1}$. }
 \item{ For $j\in\{2,\dots,k\}$, sample $s_{t,j}$ conditional on 
${s_{t-1,j}}$, where $s_{t,1:j-1}$ are removed from the support of the discrete uniform if need be. }
\end{enumerate}
Finally, each component $\theta_{t,j}$ of $\theta_{t,1:k+1}$ was independently propagated via a log normal distribution with mean that depended on the value of $\theta_{t-1,j}$.

\subsection{Sequential Monte Carlo (SMC) of \cite{Toni_2009}}
We also implemented the ABC-SMC algorithm for model selection on \cite[page 190]{Toni_2009}, with steps labelled MS1-MS3. Within the perturbation kernel, each $l^{\text{th}}$ component of $\theta_{j}$ was independently propagated via a log normal distribution with a mean that depended on the value of $\theta_{j}^l$ from the previous iteration of SMC. Each individual change-point was propagated via a discrete uniform random walk. If the site labels range from one to $m$, then any values of the support of the random walk that were less than two or greater than $m$ were assigned a zero probability of being chosen. Furthermore, to avoid selecting the same site twice at iteration $i$, the sampling consisted of the same two steps enumerated above for the PMMH algorithms.

\subsection{Time Machine}\label{sec:ImplementationDetailsEx1TM}
In belief propagation \cite{Pearl_1982}, one first sends messages up from the leaves of a tree to its topmost parent nodes. When at the topmost nodes, it is required to input the marginal of each individual parent node and then send the messages back down to the leaves (i.e., one must input the probability that a parent node is of a particular type at that point in time). We approximated those probabilities with the model's equilibrium frequencies when employing the time machine.

%We used a Dirichlet multinomial distribution (for an ordered sample) to approximate the stationary distribution of the tree. The parameters of this distribution are
%\begin{enumerate}
% \item{ $\{ \nu_i \}_{i\in\{1,\dots,p\}}$, where each $\nu_i$ is the probability that a node is of type $i$, and }
% \item{ $\Sigma$, which is a scalar that multiplies each $\nu_i$. }
%\end{enumerate}
%In our numerical examples, we set $\Sigma=1$ and $\{ \nu_i \}_{i\in\{1,\dots,p\}}$ to be the known equilibrium frequencies of the underlying DNA model of evolution.

\subsection{Summary Statistic}\label{sec:ImplementationDetailsEx1sumstat}
The two ABC algorithms simulated datasets given sampled values of $k$, $s_{1:k}$, and $\theta_{1:k+1}$. These model parameters were accepted as output when the simulated data was deemed to be sufficiently close to the actual dataset. Mean pairwise difference is sometimes used in ABC algorithms to compare two genetic sequences (see \cite[Section 2.6]{Lopes_2009}). Our datasets had $n\geq 1$ genetic sequences each. When determining if simulated data was ``sufficiently close'' to an actual dataset, we compared each of the $n$ genetic sequences from the simulated data to its counterpart in the actual dataset. If the sum of the sites with different values (across all $n$ genetic sequences) was less than a specified tolerance level, then the simulated and actual datasets were deemed to be sufficiently close to one another.

In both of the ABC algorithms implemented here, the terminal tolerance level at step $T$ of the SMC algorithm was $(mn/3)$. This value is as low as we were able to set the tolerance level. When we tried reducing the level further, the PMMH which employed ABC accepted almost no MCMC moves and the ABC-SMC of \cite{Toni_2009} was unable to accept any samples at step $T$ (regardless of how high we set $T$ or $N$ to facilitate jumping between temperatures). However, with the tolerance level set as such, the ABC-SMC of \cite{Toni_2009} was very fast, and it produced 5,000 samples from all of the tested models within a matter of minutes. We tried increasing $N$ so that the algorithm could use the full six hours alloted, but even when we set $N$ to be a very high number, the resulting inference did not change at all and was consistently extremely poor.

\subsection{Parallelisation}
The algorithms were implemented in C++, using the OpenMP 3.0 and Eigen \cite{Guennebaud_2010} libraries. The computation of the likelihood was parallelised across $m$ in the PMMH algorithms which employed the time machine. In the ABC algorithms, simulation of the sites of the sequences was parallelised across $m$.

\section{Supporting Information: Implementation for Section 4.2}\label{sec:ImplementationDetailsEx2}
Our particle marginal Metropolis-Hastings (PMMH) algorithm \cite{Andrieu_2010} which employed the time machine used $N=50$ particles, and the sequential Monte Carlo (SMC) sampler \cite{DelMoral_2006} within this algorithm ran for $T=150$ time steps. The unknown model parameters were propagated from PMMH iteration $(i-1)$ to iteration $i$ using the same schemes as outlined in Section \ref{sec:ImplementationDetailsEx1pmmh}. For the time machine approximation to the stationary distribution of the tree, we used the same scheme as stated in Section \ref{sec:ImplementationDetailsEx1TM}. Finally, the algorithm was implemented in C++, using the OpenMP 3.0 and Eigen \cite{Guennebaud_2010} libraries, and the computation of the likelihood was again parallelised across $m$.

We set $g=4$, which means that the leftmost three nodes were removed from the tree in Figure 1 of the main paper. We recognize that this is a deep cut for such a small tree and that the accuracy of the algorithm suffered as a result. However, the goal of this exercise was to get a sense if the algorithm could be used on real data and not necessarily to procure very accurate estimates of the true model parameters. A deep cut was necessary to facilitate mixing and enable the algorithm to complete a large number of iterations quickly for this test.

\end{document}